\newcommand{\bra}[1]{\langle #1|}
\newcommand{\ket}[1]{|#1\rangle}

\RequirePackage{fix-cm}
\documentclass[smallextended]{svjour3}       
\smartqed  
\usepackage{graphicx}
\usepackage{graphicx}
\usepackage{amsmath}
\usepackage{epstopdf}
\usepackage{comment}

\usepackage{cite}
\usepackage{amsmath,amssymb,amsfonts}
\usepackage{algorithmic}
\usepackage{graphicx}
\usepackage{textcomp}
\usepackage{mathrsfs}
\usepackage{amssymb}
\usepackage{graphicx}
\usepackage{amsmath}
\usepackage{algorithmic}
\usepackage{graphicx}
\usepackage{graphics}
\usepackage{url}
\usepackage{blkarray}
\usepackage{mathtools}
\usepackage{float}
\usepackage{multicol}
\usepackage{cite}
\usepackage{graphicx}
\usepackage{epstopdf}
\usepackage{epsfig}
\usepackage{stmaryrd}
\usepackage{gensymb}
\usepackage[utf8]{inputenc}
\usepackage{color}
\usepackage{multirow}
\usepackage[ruled,vlined]{algorithm2e}
\IncMargin{-\parindent}
\DeclarePairedDelimiter{\ceil}{\lceil}{\rceil}

\begin{document}

\title{Circuit Design for $k$-coloring Problem and Its Implementation in Any Dimensional Quantum System
}


\author{Amit Saha         \and
        Debasri Saha  \and
        Amlan Chakrabarti 
}


\institute{Amit Saha \at
              A. K. Choudhury School of Information Technology,  University of Calcutta, Kolkata - 700 106, India
              \email{abamitsaha@gmail.com}           
           \and
           Debasri Saha \at
              A. K. Choudhury School of Information Technology,  University of Calcutta, Kolkata - 700 106, India
              \and
           Amlan Chakrabarti \at
              A. K. Choudhury School of Information Technology,  University of Calcutta, Kolkata - 700 106, India
}

\date{Received: date / Accepted: date}

\maketitle

\begin{abstract}
With the evolution of quantum computing, researchers now-a-days tend to incline to find solutions to NP-complete problems by using quantum algorithms in order to gain asymptotic advantage.  In this paper, we solve $k$-coloring problem (NP-complete problem) using Grover's algorithm in any dimensional quantum system or any $d$-ary quantum system for the first time to the best of our knowledge, where $d \ge 2$. A newly proposed comparator-based approach helps to generalize the implementation of the $k$-coloring problem in any dimensional quantum system. Till date, $k$-coloring problem has been implemented only in binary and ternary quantum system, hence, we abide to $d=2$ or $d=3$, that is for binary and ternary quantum system for comparing our proposed work with the state-of-the-art techniques. This proposed approach makes the reduction of the qubit cost possible, compared to the state-of-the-art binary quantum systems. Further, with the help of newly proposed ternary comparator,  a substantial reduction in quantum gate count for the ternary oracle circuit of the $k$-coloring problem than the previous approaches has been obtained. An end-to-end automated framework has been put forward for implementing the $k$-coloring problem for any undirected  and unweighted graph on any available Near-term quantum devices or Noisy Intermediate-Scale Quantum (NISQ) devices or multi-valued quantum simulator, which helps in generalizing our approach.  
\end{abstract}

\section{Introduction}
\label{intro}
Modern day researchers has shown a startling interest for implementing quantum algorithms \cite{chuang, qc, shor}, which give a potential speedup over many of their classical counterparts as the advancement of quantum computer has achieved a phenomenal success in recent years \cite{preskil}. There is an immense urge for the implementation of NP-complete problems on quantum computers with the thriving quantum wave \cite{arpitadi}. According to the seminal work on computational complexity by Karp \cite{karp}, if a solution to any of the NP-complete problems can be obtained then any other NP-complete problem can be polynomially reducible to that problem. In this paper, we have focused on one of the well-known NP-complete problems i.e., $k$-coloring problem. Our main focus is to provide an end-to-end framework that automatically implements an NP-complete problem i.e., $k$-coloring problem in any dimensional quantum system, so that if anyone can map their computational problem to the $k$-coloring problem in polynomial time, will be able to implement further, without prior knowledge of gate-based quantum circuit implementation.    

An automatic circuit synthesis for $k$-coloring problem with the help of Grover's
algorithm \cite{grover} is presented in the context of multi-valued quantum
system in this paper. We graduate to multi-valued quantum system or qudits \cite{muthu}, which in the course reduce the circuit complexity and commend the efficiency of quantum algorithms \cite{amit_IEEEtran} to provide larger state space with simultaneous multiple control operations \cite{ternary_algo, zheng_ternary, bocharov}. For instance, $N$ qubits can be formulated as $\frac{N}{log_2 {d}}$ qudits, which immediately gives $log_2 {d}$-factor in run-time \cite{Fan02, khan_multi, qudit_graph}. The multi-valued quantum system can be realized on mostly available quantum technologies, as for example, ion trap \cite{31}, continuous spin systems \cite{1,4}, topological quantum systems \cite{6, 13}, nuclear magnetic resonance \cite{17}, photonic systems \cite{19},  superconducting transmon technology \cite{32} and molecular magnets \cite{34}.
   
\par  $k$-coloring problem is an NP-complete problem that assigns colors to every nodes or vertices of a given graph with the available $k$ colors in such a way that every adjacent vertices connected by an edge having distinct colors. Suppose $n$ is the number of nodes of a given graph, $k$ is the number of colors, then to find an appropriate solution using a classical algorithm requires $O(2^{n*log_{d} k})$ number of steps in $d$-dimensional system \cite{amit_completecolor, sbm_graphcolor}. On the other hand, by using oracle and the diffusion operator of multi-valued  Grover's algorithm \cite{quantumsearch_qudits, grover_qudit}, finding the exact solution needs $O\sqrt{N}$ number of iterations where $N$ is $2^{n*log_{d} k}$. 

Many researchers have already addressed graph coloring problem for binary and ternary quantum system. Earlier in \cite{amit_completecolor, graph_color}, graph coloring problem with the help of Grover's algorithm is talked about with respect to the binary quantum systems. Again, in \cite{ibm_graphcolor}, SAT reduction technique, the state-of-the-art approach, is used for solving 3-coloring problem and it gives an end-to-end framework for the implementation of it in the IBMQ quantum processor \cite{IBM}. But,  SAT reduction technique generates an immense qubit cost, resulting in inefficient circuit cost. Previously in \cite{wang_color} and \cite{sbm_graphcolor}, circuit synthesis for graph coloring problem using Grover’s algorithm has been presented with respect to ternary quantum system with the help of ternary comparators, which have been proposed in \cite{ternary_comparator}. Albeit, in these works, the gate cost remains colossal.

 In \cite{amit_kcolor},  we proposed a comparator-based approach for implementing the $k$-coloring problem in binary quantum structure, which has less qubit cost as compared to the state-of-the-art. In this paper, we have generalized the comparator for $d$-ary quantum system, which helps to overcome the engineering challenge of the implementation of $k$-coloring problem in $d$-ary quantum system. We have proposed an automated end-to-end framework for any dimensional quantum system to implement $k$-coloring problem using newly proposed generalized comparator to map the high level description of proposed circuit to any hardware-level quantum operations with an abstraction with better quantum cost in terms of quantum gate cost relative to the state-of-the-art works. In addition to this, we have claimed the following, for further establishing the novelty of the propose research work: 
\begin{itemize}
    \item We propose an automated end-to-end framework for $k$-coloring problem using quantum search algorithm in any dimensional quantum system for the first time, to the best of our knowledge. 
   
    \item The design of the proposed framework is such that, the quantum solution of $k$-coloring problem can be mapped into any available near-term quantum devices/multi-valued quantum technology, which makes our approach generalized in nature.
    
    \item We show that our newly proposed comparator helps to implement $k$-coloring problem with reduced quantum cost with respect to qubits and quantum gates as compared to state-of-the art approaches in binary and ternary quantum domain.
    \item A generalized comparator for $d$-ary system is elaborated in this paper, which is a first of its kind approach.
    
\end{itemize}

\par The paper is structured as follows. In section 2, the brief description of Grover's algorithm and quantum circuits are described. The proposed methodology is vividly explained in section 3. In section 4, implementation of $k$-coloring problem has been shown. Concluding remarks are captured in Section 5.

\section{Background}

In this section, we have explicitly described quantum circuit and Grover's algorithm.

\subsection{Quantum circuit}
Any quantum algorithm can be expressed or visualized in the form of a quantum circuit. Logical qubits/qudits and quantum gates comprise these quantum circuits \cite{Di02}. The number of gates present in a circuit is called gate count and the number of qubits/qudits present in a circuit known as qubit/qudit cost. 
\subsubsection{Qubits/Qudits}
 Logical qubit/qudit that encodes input/output of a quantum algorithm is referred to as data qubit or qudit. Ancilla qubit/qudit are another type of qubit/qudit used to store temporary results. In $d$-dimensional quantum system \textit{qudit} is the unit of quantum information. Qudit states can be manifested as a
vector in the~$d$ dimensional Hilbert space~$\mathscr{H}_d$.
 The span of orthonormal basis vectors $\{\ket0,\ket1,\ket2,\dots \ket{d-1}\}$ is the vector space. 
In qudit system, the general form of quantum state  can be expressed as 
\begin{equation}
\ket{\psi}=\alpha_0 \ket0 +\alpha_1 \ket1 +\alpha_2 \ket2+\cdots+\alpha_{d-1} \ket{d-1}=
\begin{pmatrix}
\alpha_0 \\
\alpha_1 \\
\alpha_2 \\
\vdots   \\
\alpha_{d-1} \\
\end{pmatrix}
\end{equation}
where $|\alpha_0|^2+|\alpha_1|^2+|\alpha_2|^2+\cdots+|\alpha_{d-1}|^2=1$ and $\alpha_0$, $\alpha_1$, $\dots$, $\alpha_{d-1} \in\mathbb{C}^d$.
\subsubsection{Generalized Quantum Gates}

 In this section, an outline of generalized qudit gates is conferred. The generalisation can be delineated as discrete quantum states of any arity.  In a quantum algorithm, for modification of the quamtum state, unitary qudit gates are applied. For logic synthesis of Grover's algorithm in $d$-dimensional quantum system, it is necessary to take into account one-qudit generalized gates viz. NOT gate ($X_d$), phase-shift gate ($Z_d$),  Hadamard  gate ($F_d$), two-qudit generalized CNOT gate ($C_{X,d}$) and Generalized multi-controlled Toffoli gate ($C^{n}_{X,d}$). These gates are expressed in detail for better understanding:

\textbf{Generalized NOT Gate:} $X_d$, the generalized NOT or increment gate,  for a $(d \times d)$ matrix is as follows:

 \begin{align*}
X_d = \left(\begin{matrix} 0 & 0 & \ldots & 0 & 1 \\ 1 & 0 & \ldots & 0 & 0 \\ 0 & 1 & \ldots & 0 & 0 \\ \vdots & \vdots & \ddots & \vdots & \vdots \\ 0 & 0 & \ldots & 1 & 0\end{matrix}\right)
  \end{align*}

\textbf{Generalized Phase-Shift Gate:} $Z_d$, the generalized phase-shift gate of a $(d \times d)$ matrix is as follows, with $\omega=e^{\frac{2\pi i}{d}}$;

  \begin{align*}
Z_d = \left(\begin{matrix} 1 & 0 & 0 & \ldots & 0 \\ 0 & \omega & 0 & \ldots & 0 \\ 0 & 0 & \omega^2 & \ldots & 0 \\ \vdots & \vdots & \vdots & \ddots & \vdots \\ 0 & 0 & 0 & \ldots & \omega^{d-1}\end{matrix}\right)
  \end{align*}

\textbf{Generalized Hadamard Gate:} $F_d$, the generalized quantum Fourier transform or generalized Hadamard gate, produces the superposition of the input basis states. The $(d \times d)$ matrix representation of it is as shown below  
:

\begin{align*}
F_d = {1\over\sqrt{d}} \left(\begin{matrix} 1 & 1 & 1 & \ldots & 1 \\ 1 & \omega & \omega^2 & \ldots & \omega^{d-1}  \\ 1 & \omega^2 & \omega^4 & \ldots & \omega^{2(d-1)}  \\ \vdots & \vdots & \vdots & \ddots & \vdots \\ 1 & \omega^{d-1} & \omega^{2(d-1)} & \ldots & \omega^{(d-1)(d-1)} \end{matrix}\right)
  \end{align*}

\textbf{Generalized CNOT Gate:} Quantum entanglement is a unparalleled property of quantum mechanics, and can be attained by a controlled NOT (CNOT) gate in a binary quantum system. For $d$-dimensional quantum systems, the binary 2-qubit CNOT gate is generalised to the $INCREMENT$ gate:\\ $\text{INCREMENT}\ket{x}\ket{y}=\ket{x}\ket{(x+y) \mod d}$, if $x=d-1$, and = $\ket{x}\ket{y}$, otherwise. \\ The $(d^2 \times d^2)$  matrix representation of the generalized CNOT $C_{X,d}$ gate is as follows:

\begin{equation*}
C_{X,d} = \left( \begin{matrix}
    I_d & 0_d & 0_d & \ldots & 0_d \\
    0_d & I_d & 0_d & \ldots & 0_d \\
    0_d & 0_d & I_d & \ldots & 0_d \\
    \vdots & \vdots & \vdots & \ddots & \vdots \\
    0_d & 0_d & 0_d & \ldots &  X_d \\
\end{matrix} \right)
  \end{equation*}

where $I_d$ and $0_d$ are both $d \times d$ matrices as shown below:
\begin{equation*}
I_d = 
\begin{pmatrix}
    1 & 0 & 0 & \ldots & 0 \\
    0 & 1 & 0 & \ldots & 0 \\
    0 & 0 & 1 & \ldots & 0 \\
    \vdots & \vdots & \vdots & \ddots & \vdots \\
    0 & 0 & 0 & \ldots &  1 \\
\end{pmatrix} 
\quad\textrm{and,}\quad
0_d =  
\begin{pmatrix}
    0 & 0 & 0 & \ldots & 0 \\
    0 & 0 & 0 & \ldots & 0 \\
    0 & 0 & 0 & \ldots & 0 \\
    \vdots & \vdots & \vdots & \ddots & \vdots \\
    0 & 0 & 0 & \ldots &  0 \\
\end{pmatrix}
\end{equation*}

\textbf{Generalized Multi-controlled Toffoli Gate:} We extend the generalized CNOT or $INCREMENT$ further to operate over $n$ qudits as a generalized Multi-controlled Toffoli Gate or $n$-qudit Toffoli gate $C_{X,d}^n$.  For $C_{X,d}^n$, the target qudit is increased by $1 \ (\text{mod } d)$ only when all $n-1$ control qudits have the value $d-1$. The $(d^n \times d^n)$ matrix representation of generalized Multi-controlled Toffoli (MCT) gate is as follows:

\begin{equation*}
C_{X,d}^n = \left( \begin{matrix}
    I_d & 0_d & 0_d & \ldots & 0_d \\
    0_d & I_d & 0_d & \ldots & 0_d \\
    0_d & 0_d & I_d & \ldots & 0_d \\
    \vdots & \vdots & \vdots & \ddots & \vdots \\
    0_d & 0_d & 0_d & \ldots &  X_d \\
\end{matrix} \right)
  \end{equation*}

Owing to technology constraints, a multi-controlled Toffoli gate can be substituted by an equivalent circuit comprising one-qudit and/ two-qudit gates, although at first the multi-controlled Toffoli must be decomposed into a set of Toffoli gates for any dimensional quantum system.

\begin{figure}[!htb]
 \centering
 \includegraphics[width=120mm,height=60mm]{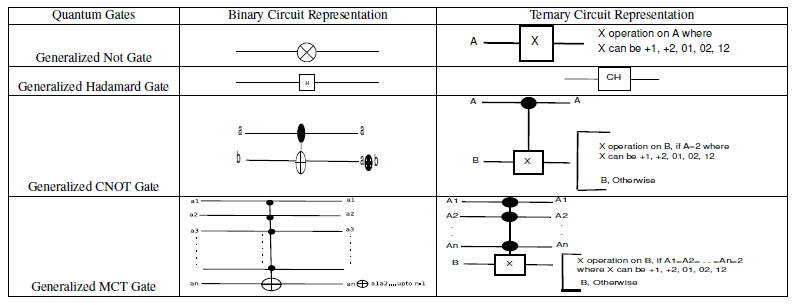}
 \caption{Circuit Representation of Generalized Quantum Gates}
 \label{table1}
 \end{figure}

The binary \cite{shor02} and ternary circuit \cite{muthu, Di} representation of  all the above mentioned generalized gates are described in Figure \ref{table1}.

\subsection{Generalized Grover's Algorithm in $d$-dimensional Quantum System}\label{sec3}

The generalized Grover's algorithm in $d$-dimensional quantum system is presented here. There exists two sub-parts of the algorithm viz. oracle and diffusion \cite{grover_qudit}. As per convention, Grover's algorithm for searching in an unstructured database can be defined as follows: given a collection of unstructured database elements $x=1, 2 ,\ldots, N$, and an \emph{oracle function} $f(x)$ that acts on a marked element $s$ as follows \cite{quantumsearch_qudits},
\begin{equation}
f(x) =\left\{\begin{array}{ll}
1,& x=s, \\
0,& x\neq s,
\end{array}\right.
\label{oracle_fun}
\end{equation}
perceive the marked element with as few calls to $f(x)$ as possible \cite{quantumsearch_qudits, grover_qudit}. The database is encoded into a superposition of quantum states where each element is assigned to a corresponding basis state. Grover's algorithm searches over every possible outcome, which is put forward as a basis vector $\ket{x}$ in an $n$-dimensional Hilbert space in $d$-dimensional quantum system. Likewise, the marked element is encoded as $\ket{s}$. Thus, the search can be done in parallel, after application of unitary operations as an oracle function to the \emph{superposition} of the different possible outcomes. The generalized diffusion operator, also known as inversion about the average operator, amplifies the amplitude of the marked state to increase its measurement probability using constructive interference, with simultaneous enfeeblement of all other amplitudes, and searches the marked element in $O(\sqrt{N})$ steps, where $N=d^n$ \cite{quantumsearch_qudits}.

The circuit diagram for the generalized Grover's algorithm in a $d$-dimensional quantum system is presented in Figure \ref{grover}, where at least $n + 1$ qudits are required. More elaborately, the steps of the Grover's algorithm are as follows:
 \begin{figure}[!htb]
 \centering
 \includegraphics[width=90mm,height=2.2cm]{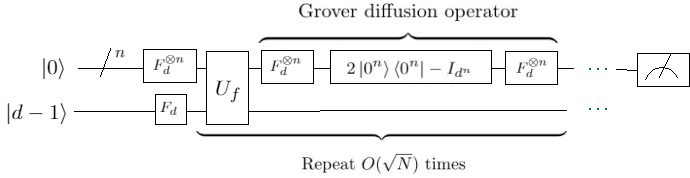}
 \caption{Generalized Circuit for Grover's algorithm in $d$-dimensional quantum system}
 \label{grover}
 \end{figure}

\par \textbf{Initialization}: The algorithm starts with the uniform superposition of all the basis states on the $n$ input qudits in $\ket{0}$ by assimilating generalized Hadamard or quantum DFT gate. The last ancilla qudit is used as an output qudit which is initialized to $F_d \ket{d-1}$. Thus, we obtain the $d$-dimensional quantum state $\ket{a}$:
\[\vert a\rangle=F_d^{\otimes{n}}\vert 0_d\rangle=\,{1\over\sqrt{d^{n}}}\sum_{x=1}^{d^{n}}\vert x\rangle\]

\textbf{Oracle query}: The oracle ($U_f$) of Grover's search marks the marked state $\ket{s}$ while keeping all the other states unchanged, and can be expressed as:
\[\vert x\rangle\xrightarrow{U_f}(-1)^{f(x)}\vert x\rangle \]

The oracle block $U_f$ as shown in Figure \ref{grover} is dependant on the problem instance. The oracle using Unitary transformation is needed to be designed as per requirement.

\textbf{Diffusion}: The diffusion operator of Grover's search is generic, it doesn't depend on specific problem . As shown in Figure \ref{grover}, the diffusion operator is initially assigned with generalized Hadamard $(F_d^{\otimes{n}})$, then by  $2\ket{0^n}\bra{0^n}-I_{d^n}$ and generalized Hadamard $(F_d^{\otimes{n}})$ again. The diffusion operator $(D)$ can be expressed as:
\[D=F_d^{\otimes{n}}  [2\ket{0^n}\bra{0^n}-I_{d^n}] F_d^{\otimes{n}}\]

The matrix representation of generalized diffusion  operator \cite{quantumsearch_qudits} for $d$-dimensional quantum system is shown below:


\begin{equation*}
diff_d = \left( \begin{matrix}
    \frac{2}{d^n}-1 & \frac{2}{d^n} & \frac{2}{d^n} & \ldots & \frac{2}{d^n} \\
    \frac{2}{d^n} & \frac{2}{d^n}-1 & \frac{2}{d^n} & \ldots & \frac{2}{d^n} \\
    \frac{2}{d^n} & \frac{2}{d^n} & \frac{2}{d^n}-1 & \ldots & \frac{2}{d^n} \\
    \vdots & \vdots & \vdots & \ddots & \vdots \\
    \frac{2}{d^n} & \frac{2}{d^n} & \frac{2}{d^n} & \ldots &  \frac{2}{d^n}-1 \\
\end{matrix} \right)
  \end{equation*}

The combination of the oracle and the diffusion gives generalized Grover operator $G$,
\[G=D  U_f\]

We need to din the Grover's operator $O(\sqrt{N})$ times to get the coefficient of the marked state $\ket{s}$ large enough so that it can be obtained from measurement with probability close to 1 and thus round off the Grover's algorithm.

\section{Proposed Methodology of Circuit Synthesis for $k$-coloring Problem using Grover's Algorithm in any Dimensional Quantum System}

\begin{figure}[!h]
\centering
\includegraphics[width=150mm, height =18 cm]{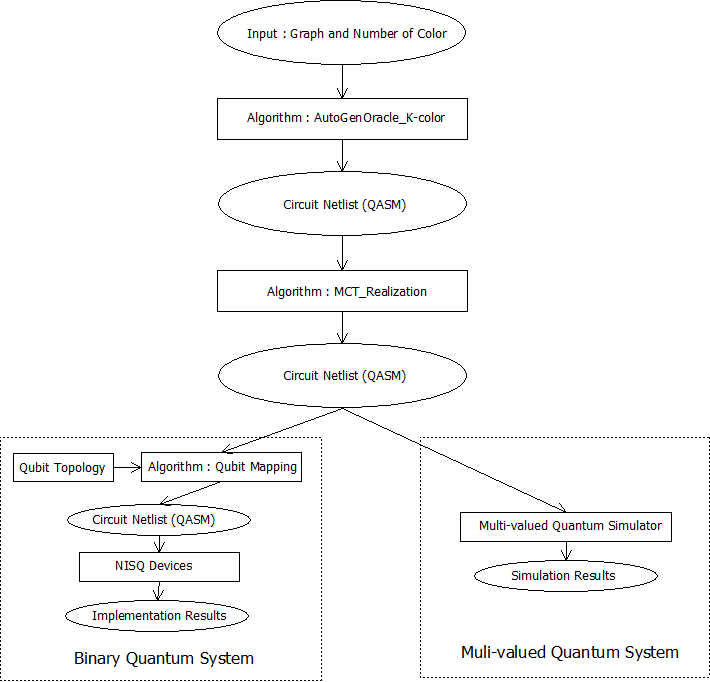}
\caption{Flowchart of our proposed work} 
\label{flow}
\end{figure}

The flowchart in Figure \ref{flow} depicts the complete flow of our proposed automated end-to-end framework. Mainly three algorithms: AutoGenOracle\_K-color, MCT\_Realization and Qubit Mapping (Binary) are the basis of our framework. At first, AutoGenOracle\_K-color algorithm, which is based on the newly designed comparator, is implemented with inputs: graph information i.e, adjacency matrix of the given graph and the number of colors ($k$) to get the output, quantum circuit netlist in the form QASM. Now, MCT\_Realization algorithm takes as input, generated circuit netlist   and  MCT gates to NISQ hardware compatible 1-qubit and 2-qubit gates \cite{portugal} and multi-valued quantum technology compatible 1-qudit and 2-qudit gates are realised. Due to multi-valued quantum technology constraint, implementation of multi-valued circuit netlist on multi-valued quantum hardware is technologically infeasible till date, rather we have simulated and verified the multi-valued circuit netlist on multi-valued simulator. Finally, qubit mapping \cite{sabre, olsq} algorithm has been used for mapping generated circuit to the NISQ devices based on the qubit topology.

The proposed methodology for the oracle circuit synthesis of the $k$-coloring problem in any dimensional quantum system as an application of the Grover's search algorithm is sketched in this section. 

\subsection{Proposed Oracle for $k$-coloring Problem in $d$-dimensional Quantum System}
\begin{figure*}[!h]
\centering
\includegraphics[width=5in, height=2in]{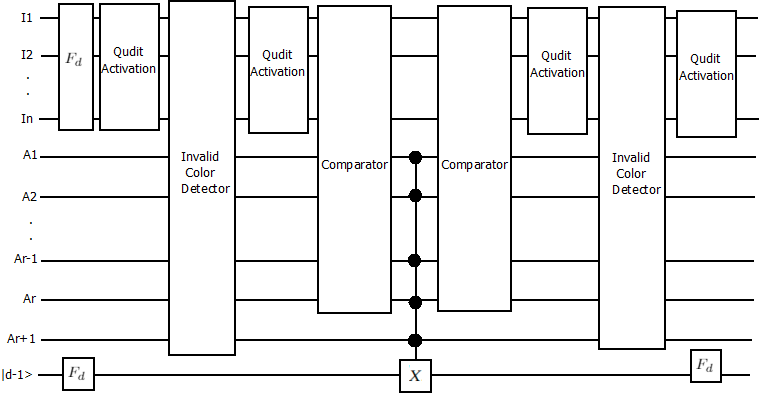}
\caption{Block Diagram of Generalized Oracular Circuit}
\label{gen}
\end{figure*}

 Figure \ref{gen} shows the quantum circuit block of oracle for the $k$-coloring problem in $d$-dimensional quantum system. The construction of oracle for $k$-coloring problem is divided into four parts. It starts with initialization, which is essentially required in Grover's Algorithm.

\subsubsection{Initialization}
Let there be a graph which has $n$ vertices and $e$ edges, $k$ is the given number of colors, then the total number of data qudits that are required for representing all the colored vertices are $n * \ceil{\log_d k}$. The oracle performs a check to find all possible right combination of properly colored vertices with $k$/fewer colors from a combination of all possible colored vertices. A superposition of $m=n*\ceil{\log_d k}$ qudits thus generates all possible combination of colored vertices. The initial data qudits in Figure \ref{gen} include $m$ qudits prepared in the ground state $\vert\psi\rangle=\ket{0}^{\otimes m}$, due to the re-usability property of ancilla qudits, $r= n$ ancilla qudits in the exited state $\ket{\theta}=\ket{d-1}^{\otimes r}$ (These $r$ ancilla qudits are required to prepare Invalid\_Color detector block and comparator block that are thoroughly explained in the next subsection), one ancilla qudit in the ground state $\ket{\zeta}=\ket{0}$ (1 ancilla is required if invalid color exists), and one output qudit in the excited state $\ket{\phi}=\ket{d-1}$ is required to perform generalized CNOT/Toffoli/MCT operation of the oracle. This entire initialization can be mathematically depicted as:

\[\vert\psi\rangle\otimes\vert\theta\rangle\otimes\vert\zeta\rangle\otimes\vert\phi\rangle=\vert 0\rangle^{\otimes m}\otimes\vert d-1\rangle^{\otimes r}\otimes\vert 0\rangle\otimes\vert d-1\rangle\]

\subsubsection{Generalized Hadamard Transformation}

The next step after the initialization is Hadamard transformation. The genralized Hadamard transform ${F_{d}}^{\otimes m}$ on data qudits and $F_d$ on output qudit is performed, hence all possible states are superposed as $\ket{\psi_{0}}\otimes\ket{\theta_{0}}\otimes\ket{\zeta_{0}}\otimes\ket{\phi_{0}}$, where

\[\vert\psi_{0}\rangle=\,{1\over\sqrt{d^{m}}}\sum_{i=0}^{d^{m}-1}\vert i\rangle\]

\[\ket{\theta_{0}}=\ket{(d-1)(d-1)(d-1)(d-1) ..... r(times)}\]

\[\ket{\zeta_{0}}=\ket{0}\]

\[\vert\phi_{0}\rangle=\,{1\over\sqrt{d}}\left(\vert 0\rangle+\omega^{d-1}\vert 1\rangle+\omega^{2(d-1)}\vert 2\rangle+ ..... +\omega^{(d-1)(d-1)}\vert d-1\rangle\right)\]

\subsubsection{Proposed $U_f$ Transformation for d dimensional system:}

The proposed unitary $U_f$ transformation is distinctly divided into two parts.
\par \textbf{(1)Reduction of Invalid Colors:} As color, $c=\ceil{\log_d k}$, so  maximum $d^c$ colors are considered. If $d^c=k$, then all colors are valid colors, else there exists a set of $d^c -k$ invalid colors. The valid colors are used to optimize the search space. Using the following steps this can be executed: 

\par \textbf{Qudit Activation:} Colors should be numbered sequentially as $\{0,1,2 .... d^{c}-1\}$. After the generalized Hadamard transformation, the input data qudit lines act as the $d$-dimensional representation of combination of all possible colored vertices. The oracle performs a check for only $k$, which is the combination of valid colors. All the input qudit lines should be in the excited state $\ket{d-1}$ for those particular combinations of invalid colors by making input qudit lines suitable as control lines for Generalized CNOT/Toffoli/MCT operation to be assured that the oracle is checking only the $k$-colored combination of vertices. A number of generalized NOT gates must be imposed on the input qudit lines, that are in the ground state $\ket{0}$ accompanied by the application of 'Invalid Color Detector'.  Again after 'Invalid Color Detector', this 'Qudit Activation' is to be applied to roll back to the initial superposed quantum state.

\textbf{Invalid Color Detector:}  In any combination of colored vertices if any invalid color is noticed then that combination is rejected with the use of the function ICD (Invalid Color Detector) as:
\begin{equation}
\resizebox{.85\hsize}{!}{$ Generalized\_ICD(I_1, I_2, .., I_n, f)= \left\{\begin{array}{ll}\mbox{$f=0$,} & \mbox{if $I_1 or I_2 or.. I_n =$ Invalid color};\\
\mbox{$f=d-1$,} & \mbox{No invalid color}.\\
\end{array}\right. $}
\end{equation}

\begin{figure}[!h]
\centering
\includegraphics[width=45mm,height=3cm]{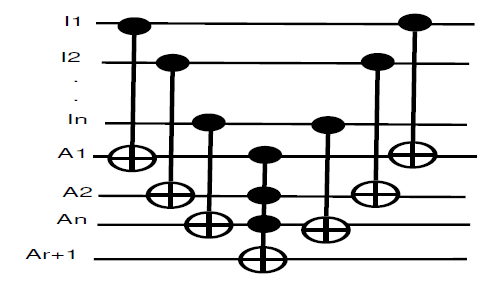}
\caption{Binary Invalid Color Detector}
\label{b_ICD}
\end{figure}

The circuit synthesis of 'Invalid Color Detector', that is functionally depicted in Equation 4 for $n$ vertices, where $I_1, I_2, .., I_n$ are the data qubits, is described in Figure \ref{b_ICD} .

\begin{equation}
\resizebox{.85\hsize}{!}{$ Binary\_ICD(I_1, I_2, .., I_n, f)= \left\{\begin{array}{ll}\mbox{$f=0$,} & \mbox{if $I_1 or I_2 or.. I_n =$ Invalid color};\\
\mbox{$f=1$,} & \mbox{No invalid color}.\\
\end{array}\right. $}
\end{equation}

\begin{figure}[!h]
\centering
\includegraphics[width=50mm,height=3.5cm]{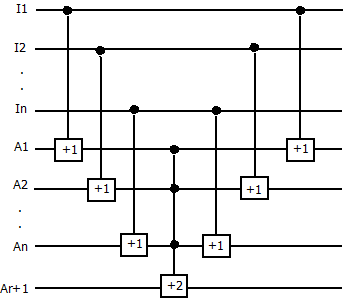}
\caption{Ternary Invalid Color Detector}
\label{ICD}
\end{figure}

 The circuit synthesis of 'Invalid Color Detector' for ternary quantum system is described in Figure \ref{ICD}. That is again, functionally shown in Equation 5 for $n$ vertices, where $I_1, I_2, .., I_n$ are the data qutrits.

\begin{equation}
\resizebox{.85\hsize}{!}{$ Ternary\_ICD(I_1, I_2, .., I_n, f)= \left\{\begin{array}{ll}\mbox{$f=0$,} & \mbox{if $I_1 or I_2 or.. I_n =$ Invalid color};\\
\mbox{$f=2$,} & \mbox{No invalid color}.\\
\end{array}\right. $}
\end{equation}

\textbf{(2)Comparator:} A newly proposed generalized comparator circuit for $d$-dimensional quantum system can be defined as:

\begin{equation}
\resizebox{.55\hsize}{!}{$ Generalized\_Comparator(a, b, f)= \left\{\begin{array}{ll}\mbox{$f=0$,} & \mbox{if $a = b$};\\
\mbox{$f=d-1$,} & \mbox{$a \neq b$}.\\
\end{array}\right. $}
\end{equation}

where $a$ and $b$ are the comparing inputs representing the colored vertices of the given graph and the ancilla qudit is $f$ .

\begin{figure}[!h]
\centering
\includegraphics[width=80mm,height=3cm]{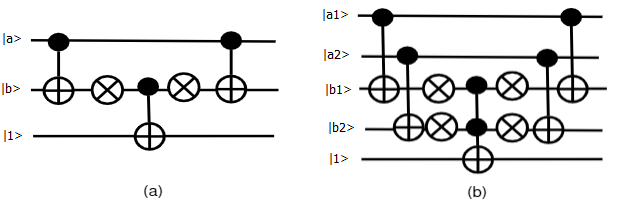}
\caption{Example Binary Comparator: (a) 2-qubit; (b) 4-qubit}
\label{comparator}
\end{figure}

In Figure \ref{comparator}, circuit synthesis for 2-qubit and 4-qubit  comparator is shown,  the functional description of which is given in Equation 7. The complete circuit synthesis for the binary comparator is designed using  CNOT, NOT, Toffoli/MCT gates. 

\begin{equation}
\resizebox{.55\hsize}{!}{$ Binary\_Comparator(a, b, f)= \left\{\begin{array}{ll}\mbox{$f=0$,} & \mbox{if $a = b$};\\
\mbox{$f=1$,} & \mbox{$a \neq b$}.\\
\end{array}\right. $}
\end{equation}

Induced from the above, a newly proposed ternary comparator can be defined as:

\begin{equation}
\resizebox{.55\hsize}{!}{$ Ternary\_Comparator(a, b, f)= \left\{\begin{array}{ll}\mbox{$f=0$,} & \mbox{if $a = b$};\\
\mbox{$f=2$,} & \mbox{$a \neq b$}.\\
\end{array}\right. $}
\end{equation}

\begin{figure}[!htb]
\centering
\includegraphics[width=55mm,height=2.0cm]{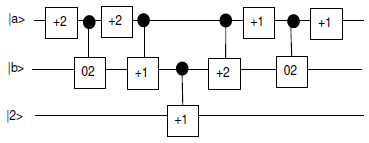}
\caption{Ternary Comaparator}
\label{Fig5}
\end{figure}

Circuit synthesis for $2$-qutrit ternary comparator is shown
in Figure \ref{Fig5}. M-S gate and 1-qutrit permutative gate
are used to design the complete circuit synthesis for ternary comparator. Initially, the total number of gate count of our newly proposed comparator is reduced by one as compared to the comparator proposed in \cite{ternary_comparator}. But, gate count can be reduced further at the time of oracular circuit synthesis as one of the outputs of our comparator is $'2'$ instead of $'1'$ \cite{ternary_comparator}.  

\begin{figure}[!htb]
\centering
\includegraphics[width=55mm,height=2.0cm]{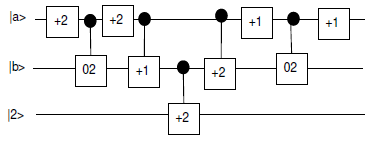}
\caption{Ternary Inverse Comaparator}
\label{Fig6}
\end{figure}

As everything has to be mirrored at the time of circuit synthesis in
order to eliminate the cost of wires, there is a need to design an inverse comparator. Our newly proposed $2$-qutrit ternary inverse comparator is shown in Figure \ref{Fig6}. This newly proposed ternary inverse comparator has a reduced gate count compared to the work in \cite{ternary_comparator}.

\subsubsection{Generalized MCT Operation}
The  $\vert\phi_{0}\rangle$, output qudit state, is initialized as:
\[{1\over\sqrt{d}}\left(\vert 0\rangle+\omega^{d-1}\vert 1\rangle+\omega^{2(d-1)}\vert 2\rangle+ ..... +\omega^{(d-1)(d-1)}\vert d-1\rangle\right)\] Applying a Generalized  MCT gate on the output line with ancilla qudits as control, an eigenvalue kickback $\omega^{(d-1)(d-1)}$ occurs as a result. This causes a phase shift for the respective input
state(s), which in turn helps in finding out all the combinations of properly colored set of vertices.  In next subsection, we have elaborately sketched about generalized Grover's diffusion operator.

\subsection{Generalized Diffusion Operator}
The circuit implementing the function of diffusion is the second part of Grover's algorithm. The component along $\ket{\psi_{0}}$ is kept unchanged when the operation is applied to a superposition state, while inverting the components in dimensions that are perpendicular to $\ket{\psi_{0}}$. This can be represented as 

\[I_{\left\vert\psi_{0}^{\perp}\right\rangle}=-I_{\vert\psi_{0}\rangle}\]

where, \[\vert\psi_{0}\rangle=\,{1\over\sqrt{d^{n}}}\sum_{i=0}^{d^{n}-1}\vert i\rangle\]

\begin{figure}[!h]
\centering
\includegraphics[width=85mm,height=2.3cm]{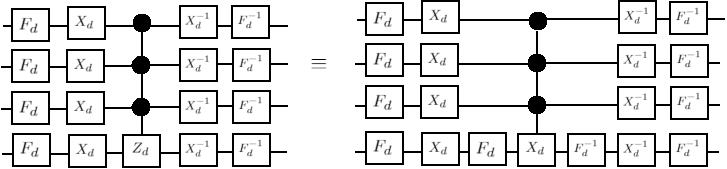}
\caption{Generalized Circuit for Grover's Diffusion Operator in $d$-dimensional quantum system \cite{quantumsearch_qudits}}
\label{diff}
\end{figure}

Figure \ref{diff} shows the generalized circuit for Grover's diffusion operator in $d$-dimensional quantum system. It can be constructed using generalized Hadamard gate, generalized NOT gate and generalized multi-controlled Toffoli gate. The algorithm that causes the gate level synthesis of the proposed method is discussed in the next subsection.

\subsection{Proposed Algorithm for Oracle Circuit Synthesis in any Dimensional Quantum System}
The proposed algorithm Algorithm 1 (AutoGenOracleK-Coloring) of automated oracular circuit synthesis for the $k$-coloring problem in any dimensional quantum system is illustrated in this subsection. The algorithm accepts the adjacency matrix of the given graph and the number of colors $k$ as input parameters.  A circuit netlist in the form of QASM is the output of the algorithm.


\begin{algorithm}[!h]
\begin{algorithmic}[1]
\label{algo:one}
\caption{AutoGenOracleK-Coloring($G(V,E)$)}\label{euclid}
\STATE \bf{INPUT} : Adjacency matrix $adj(n,n)$ of graph(G) $G(V,E)$, $V=n$ and $E=e$ where, V is the set of nodes and E is the set of edges,
Number of input data qudit lines required  $I_{r}=n*\lceil \log_{d} k \rceil$(input lines for $n$ nodes and $k$ colors, $d=d$-dimensional quantum system)$+$ ancilla lines required$= n$ $+1$ (ancila line for reduction of invalid colors(if required)) $+1$(output line($O$)),
$A_{r}$ represents ancilla line where, $1 \leq r \leq n$, $A_{r+1}$ represents ancilla line for invalid color (if required).
\STATE \bf{OUTPUT} : Circuit netlist (QASM)
\STATE Initialize $I_r$ input lines with $\ket{0}$ followed by generalized Hadamard gate ($F_d$), ancilla lines $A_r$ with $\ket{d-1}$, $A_{r+1}$ with $\ket 0$, and output line $O$ with $\ket{d-1}$ followed by a generalized Hadamard gate ($F_d$).

\STATE Apply generalized Invalid Color Detector (if required) for all possible invalid colors with suitable Qudit Activation with $I_r, A_r$ as control and $A_{r+1}$ as target.

\STATE  $l \leftarrow n$, $f \leftarrow 1$
\FOR {$i\gets1$ to $n-1$}

\STATE $r \leftarrow f$, $m \leftarrow f$
 \FOR {$j\gets{i+1}$ to $n$}
\IF {$adj(i, j) \leftarrow 1$ ($i$ and $j$ are connected by an edge($e$))}
\STATE Use a generalized comparator circuit with the input lines ($I_{i}$, $I_{j}$) corresponding $(i,j)$ as control and the ancilla line $A_{r}$  as target.
\STATE $r \leftarrow r+1$

\ENDIF
\ENDFOR
\IF {$r > f+1$}
\STATE Use a generalized Toffoli/MCT gate with all ancilla lines $A_r$ as control and $A_l$ as target
\STATE $l \leftarrow l-1$
 \FOR {$m\gets{i+1}$ to $n$}
\IF {$adj(i, j) \leftarrow 1$ ($i$ and $j$ are connected by an edge($e$))}
\STATE Use a generalized comparator circuit with the input lines ($I_{i}$, $I_{m}$) corresponding $(i,j)$ as control and the ancilla line $A_{m}$  as target.
\STATE $m \leftarrow m+1$

\ENDIF
\ENDFOR
\ELSIF {$r = f+1$ }
\STATE $f \leftarrow f+1$
\ENDIF
\ENDFOR
\STATE Use a generalized MCT gate with all the ancilla lines $ A_{1}, A_{2}, \dots A_{r+1}$ as control and $O$ as output.

\STATE Repeat step 5-26.
\STATE Repeat step 4.

\end{algorithmic}
\end{algorithm}

The total number of qudit lines that are required for the generation of the oracle circuit can be easily evaluated from the details of the adjacency matrix and the number of given colors. All the input data qudits are initialized with $\ket 0$ followed by genaralized Hadamard, ancilla lines ($A_{r}$) are initialized with $\ket {d-1}$, ancilla line $A_{r+1}$ is initialized with $\ket 0$ and the output line is initialized with $\ket {d-1}$ followed by generalized Hadamard. First of all,  Invalid Color Detector is applied with suitable Qudit Activation (if invalid color exists) with $I_r, A_r$ as control and $A_{r+1}$ as the target. After that, between two adjacent vertices$(i,j)$, a generalized comparator circuit is used with two input lines$(i,j)$ as control and the ancilla line($A_{r}$) as output. They perform the same task for all the adjacent vertices. Following this, a generalized MCT gate is used with all the ancilla lines $A_{r}$ and $A_{r+1}$ as control and the output line as output for the flip operation of the Grover's oracle. In order to achieve the mirror of the oracle circuit, we have repeated the previous steps as shown in Algorithm 1.

\subsection{Circuit Cost Estimation}

The design of generalized  oracle for our proposed algorithm has already been described. Now, in this subsection, we furnish the circuit cost analysis of the oracular circuit in Table \ref{CC}.

\begin{table}[!h]
\centering
\caption{Circuit Cost Analysis of  Oracle}
\begin{tabular}{ |c|c|c| }
  \hline

No. of Vertex & Maximum Ancilla Required & Maximum Gate Count \\ \hline
$n$ & $O(n)$ & $O(n^2 *log_{d}n)$ \\ \hline
\end{tabular}
\label{CC}
\end{table}

\begin{figure*}[!h]
\centering
\includegraphics[width=6in,height=1.5in]{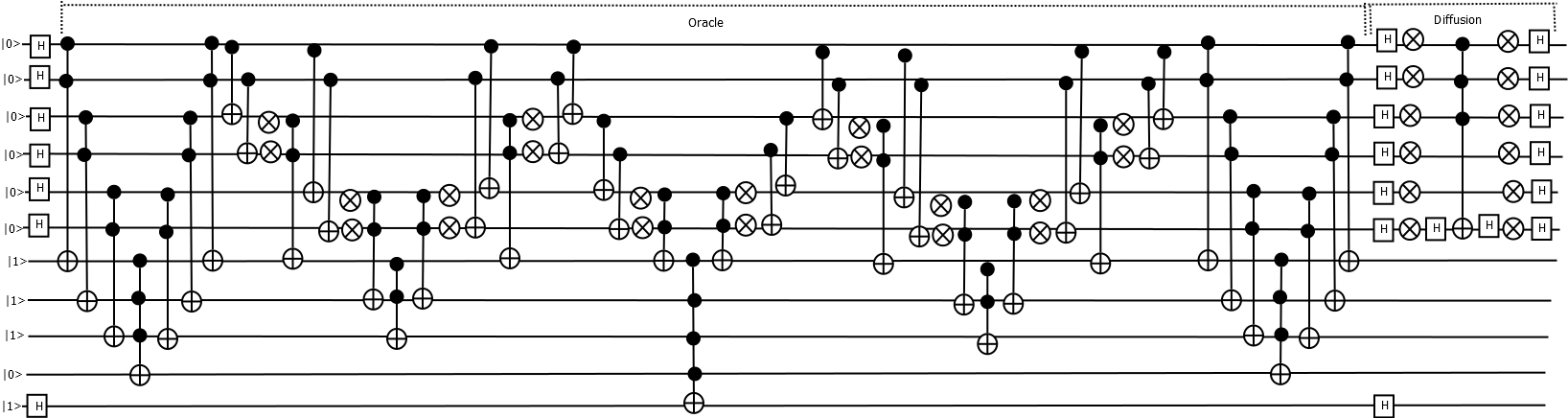}
\caption{Gate level representation of 3-coloring problem for example graph in binary quantum system}
\label{complete}
\end{figure*}

\begin{figure*}[!htb]
\centering
\includegraphics[width=150mm,height=3.5cm]{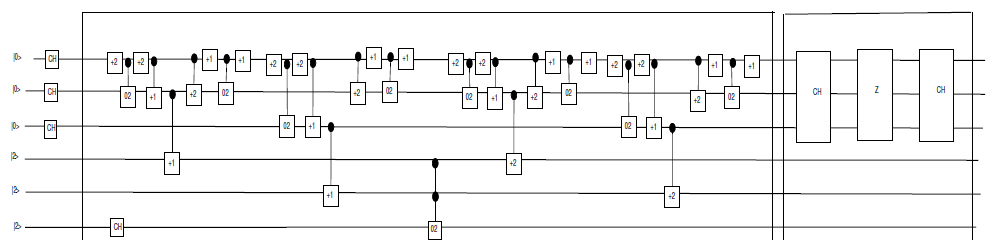}
\caption{Complete circuit for graph coloring problem of a graph with three vertices in ternary quantum system}
\label{Fig9}
\end{figure*}

  $n* \lceil log_{d} k \rceil$ data qubits are required for $n$-vertices graph and $k$ given color.  At most $n+1$ number of  ancilla and at most $O(n^2 * log_{d} n)$ gates are required to design the oracular circuit for $n$-vertices graph.  As for example, for binary quantum system, a graph of three vertices with three connected edges ($K_{3}$) is shown in Figure \ref{complete} and  a graph of three vertices with two connected edges  for ternary quantum system is shown in Figure \ref{Fig9} which illustrates the gate-optimized circuit synthesis of the 3-coloring problem,

\section{Mapping of $k$-coloring Problem to NISQ Devices/Multi-valued Quantum Simulator} 

\subsection{Mapping of $k$-coloring Problem to NISQ Devices}

\begin{figure}[!h]
\centering
\includegraphics[width=90mm, height=2.5cm]{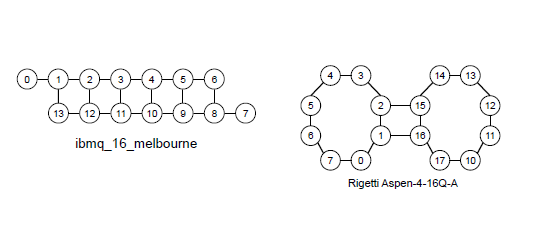}
\caption{Qubit Topology \cite{IBM, rigetti}}
\label{qt}
\end{figure}

The focus of this subsection is the mapping of generated oracle circuit to NISQ devices through MCT realization and qubit mapping algorithm. Due to the constraint of the number of qubits, \textbf{NISQ Devices} are “noisy,” thus a certain range of error should be allowed during the estimation of the simulated result of a quantum state \cite{preskil}. Superconducting quantum circuits,  quantum dot, ion trap, neutral atom are the most popular NISQ technologies for implementing the quantum circuit. For mapping the logical synthesized circuit to quantum hardware, each one of the above mentioned technologies has a specific dedicated qubit topology, as in Figure \ref{qt}. Certain 1-qubit and 2-qubit gates that are supported by most of the quantum hardware are illustrated in Table \ref{tab:gates}. The logical quantum gates are needed to be realized to hardware specific gates for making it hardware compatible for implementation.


\begin{table*}[!h]
\centering
\caption{Gate Set for NISQ Devices}
\begin{tabular}{c|c}
gate type &  gate set \\ \hline
1-qubit gates & id, x, y, z, h, r2, r4, r8, rx, ry, rz, u1, u2, u3, s, t, sdg, tdg\\
2-qubit gates & swap, srswap, iswap, xy, cx, cy, cz, ch, csrn, ms, yy, cr2, cr4, cr8, crx, cry, crz, cu1, cu2, cu3, cs, ct, csdg	
\end{tabular}
\label{tab:gates}
\end{table*}

\subsubsection{Realization of MCT Gate}

\begin{figure*}[!h]
\centering
\includegraphics[width=12cm, height=1.7in]{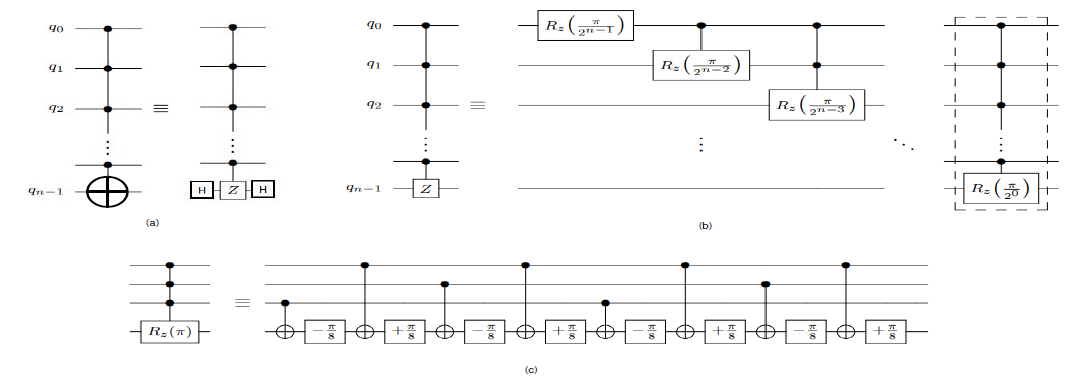}
\caption{(a) Decomposition of MCT to MCZ; (b) Decomposition of MCZ to MC$R_{x}(\pi)$ (c) Decomposition of 4-control $R_{x}(\pi)$ gate \cite{portugal}}
\label{mctreal}
\end{figure*}

The process of  decomposition of MCT gate to NISQ compatible 1-qubit and 2-qubit gates \cite{portugal} is shown in Figure \ref{mctreal}. At first, the MCT gate is to be decomposed to MCZ gate. After that, MCZ gate is realized to MC$R_{x}(\pi)$. Lastly, without using any ancilla qubit, MC$R_{x}(\pi)$ is reduced to 1-qubit and 2-qubit gates.

\subsubsection{Qubit Mapping to NISQ Devices}

 As, our proposed quantum circuit is logical, there is no constraint of qubit connectivity. There is a specific qubit topology or coupling graph for NISQ devices. The interaction between two physical qubits is defined by coupling graph. This is different for different NISQ devices. Hence, obviously  mapping the logical circuit to the physical one is challenging. The solution to this problem is to insert SWAP gates between the two qubits to comply to the hardware constraint without compromising on the logic of the quantum circuit. The concept behind a good qubit mapping problem is to minimize the number of SWAP insertion gates as well as the depth of the circuit. Li et. al. proposed SWAP-based BidiREctional heuristic search algorithm (SABRE) in \cite{sabre}, which deals with any arbitrary qubit topology for any NISQ devices. Three main features make SABRE noticeable. Firstly, an exhaustive search is not performed on the entire circuit, but a SWAP-based heuristic search is performed keeping in mind the qubit dependency. Then, initial mapping is optimized with the use of a novel reverse traversal technique. Lastly, the decay effect being introduced to enable the trade-off between the depth and the number of gates of the entire algorithm. Tan et. al. formulated layout synthesis for quantum computing, which is a benchmark, as optimization problems \cite{olsq}. They handed over two synthesizers: an
exact layout synthesizer (OLSQ) and an approximate, transition based synthesizer (TB-OLSQ). OLSQ is the first, that
guarantees optimality and efficiency both in time and
space for general quantum processors, as compared to previous exact approaches. 
This approach shows some promises of being beneficial for realistic applications for
near-term quantum computers. Our proposed circuit can easily be mapped to any arbitrary qubit topology by using one of these protocols.

  \subsection{Mapping of $k$-coloring Problem to Multi-valued Quantum Simulator} 
 \subsubsection{Realization of Generalized MCT Gate}
 
\begin{figure}[!htb]
\centering
\includegraphics[width=75mm,height=2cm]{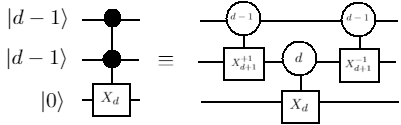}
\caption{Generalized Toffoli in $d$-dimensional quantum system \cite{amit_acmtran}}
\label{gentofdec}
\end{figure}

  In \cite{amit_acmtran}, Saha et. al. have shown the decomposition of multi-controlled Toffoli gate in $d$-dimensional quantum system. A generalized Toffoli decomposition in $d$-dimensional system using $\ket{d}$  state is shown in Figure \ref{gentofdec}. An akin construction for the Toffoli gate in binary using qutrit is evident from a previous work \cite{gokhale}; Saha et. al. have extended it for $d$-dimensional quantum system. The idea is to execute an $X_d$ operation on the target qudit (third qudit) if and only if the two control qudits, are both $\ket{d-1}$. Firstly, a $\ket{d-1}$-controlled $X^{+1}_{d+1}$, where $+1$ and $d+1$ are used to denote that the target qudit is incremented by $1 \ (\text{mod } d+1)$,  is implemented on the first and the second qudits. This eventually upgrades the second qudit to $\ket{d}$ as long as the first and the second qudits were both $\ket{d-1}$. Then, a $\ket{d}$-controlled $X_d$ gate is applied to the target qudit. Therefore, $X_d$ is executed only when both the first and the second qudits were $\ket{d-1}$, as expected. The controls are rolled back to their original states by a $\ket{d-1}$-controlled $X^{-1}_{d+1}$ gate, which reverses the effect of the first gate. The most important aspect in this decomposition is that the $\ket{d}$ state from $d+1$-dimensional quantum system can be used instead of ancilla to store temporary information.


  For further decomposition of the Toffoli for simulation purpose, the $d+1$-dimensional Toffoli gate has been decomposed into $d+2$-dimensional CNOT gates. A generalized CNOT gate for $d+2$-dimensional quantum system as $C^{+1}_{X,d+2}$, where $+1$ and $d+2$ denote that the target qudit is incremented by $1 \ (\text{mod } d+2)$ only when the control qudit value is $d+1$. The $((d+2)^2 \times (d+2)^2)$ matrix representation of the $C^{+1}_{X,d+2}$ gate is as follows:

\begin{equation*}
C^{+1}_{X,d+2} = \left( \begin{matrix}
    I_{d+2} & 0_{d+2} & 0_{d+2} & \ldots & 0_{d+2} \\
    0_{d+2} & I_{d+2} & 0_{d+2} & \ldots & 0_{d+2} \\
    0_{d+2} & 0_{d+2} & I_{d+2} & \ldots & 0_{d+2} \\
    \vdots & \vdots & \vdots & \ddots & \vdots \\
    0_{d+2} & 0_{d+2} & 0_{d+2} & \ldots  &  X^{+1}_{d+2} \\
\end{matrix} \right)
  \end{equation*}

where $X^{+1}_{d+2}$ and $0_{d+2}$ are both $(d+2) \times (d+2)$ matrices as shown below:

\begin{align*}
X^{+1}_{d+2} = 
\left(\begin{matrix} 
0 & 0 & \ldots & 0 & 1 \\ 1 & 0 & \ldots & 0 & 0 \\ 0 & 1 & \ldots & 0 & 0 \\ \vdots & \vdots & \ddots & \vdots & \vdots \\ 0 & 0 & \ldots & 1 & 0
\end{matrix}\right)
\quad\textrm{and,}\quad
0_{d+2} = 
\left(\begin{matrix} 
0 & 0 & \ldots & 0 & 0 \\ 0 & 0 & \ldots & 0 & 0 \\ 0 & 0 & \ldots & 0 & 0 \\ \vdots & \vdots & \ddots & \vdots & \vdots \\ 0 & 0 & \ldots & 0 & 0
\end{matrix}\right)
  \end{align*}

\begin{figure*}[!htb]
\centering
\includegraphics[width=142mm,height=6cm]{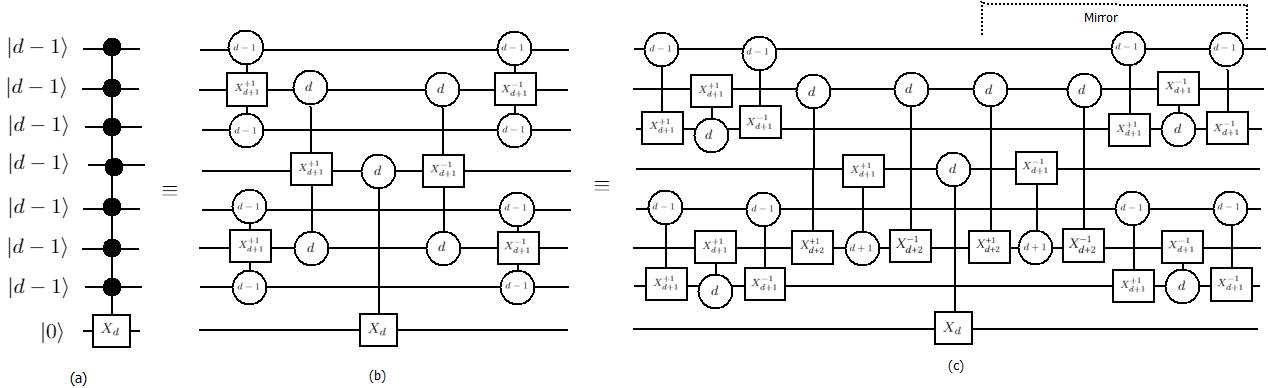}
\caption{Decomposition of 8-qudit Toffoli Gate \cite{amit_acmtran}}
\label{8quditdecompose}
\end{figure*}

As an example, a 8-qudit Toffoli gate is shown in Figure \ref{8quditdecompose}(a). First, we need to decompose it as in \cite{gokhale} as shown in Figure \ref{8quditdecompose}(b). Further, we need to decompose all the $d+1$-dimensional Toffoli gates into $(d+2)$-dimensional CNOT gates as shown in Figure \ref{8quditdecompose}(c) with the help of the Saha et. al. decomposition of the generalized Toffoli in any dimensional quantum system \cite{amit_acmtran}. All the $d-1$-controlled Toffoli gates are decomposed into $d-1$-controlled and $d$-controlled CNOT gates as shown in Figure \ref{8quditdecompose}(c). Likewise, all the $d$-controlled Toffoli gates are decomposed into $d$-controlled and $d+1$-controlled CNOT gates. Consequently, with the help of $\ket{d}$ and $\ket{d+1}$ quantum state of $(d+2)$-dimensional system, $X_d$ is caried out effectively if all the controlled qudits are in $\ket{d-1}$ state. In this way, any dimensional multi-controlled Toffoli gate can be decomposed.

\subsection{Experimental result}
\subsubsection{Experimental result of $k$-coloring Problem in NISQ Device}

 The generated oracle circuit for example graph has been taken as an example case for the simulation of $k$-coloring problem as shown in Figure \ref{complete}. The simulation is performed on IBMQ cloud based physical device \cite{4}.
 

 \begin{figure}[!h]
\centering
\includegraphics[width=120mm, height=2in]{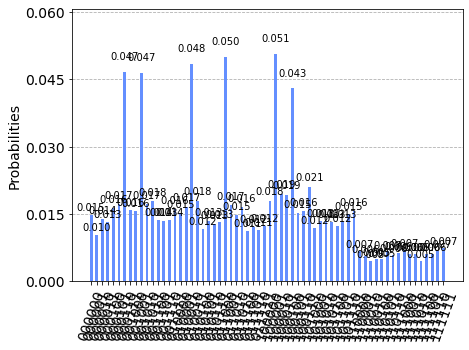}
\caption{Amplitudes of Quantum States}
\label{output}
\end{figure}

 The resultant probabilistic output after applying Grover's diffusion operator is shown in Figure \ref{output}, where the amplitude of the solution state has been amplified. The location of the solution states are $\ket{011000}$, $\ket{100100}$, $\ket{000110}$,  $\ket{010010}$, $\ket{001001}$, and $\ket{100001}$ where $00$, $01$, and $10$ are the valid colors as we take $11$ as invalid color. These are the properly colored vertex combinations in the given example graph that solves the $k$-coloring problem with high probability. As we have taken $K_3$ graph as an example graph and the three valid colors are $00$, $01$, and $10$, solution states must be all possible combination of the valid colors.

\subsubsection{Experimental result of $k$-coloring Problem in Ternary Quantum System}

Our ternary circuit instance is verified through simulation with the help of MATLAB simulator \cite{MATLAB}. Unfortunately, MATLAB has some limitations of memory constraints. For that, the verification of our circuit needs to be restricted with at most six qutrits. Therefore, a graph of three vertices is taken, where one vertex is connected with other two vertices as an example case
for simulating the graph coloring problem.  As per the Algorithm 1, as a three vertices graph is taken, three qutrits($n*\lceil \log_{3} n \rceil$) are required to represent them. As the example graph having two edges, two ancilla qutrit is required to perform the comparator circuit synthesis and one final qutrit as output.
 As three colors are given, which are encoded as $\ket{0}, \ket{1}$ and $\ket{2}$. Generally,
a graph with three vertices can be colored with the help of three colors in
$3^3=27$ different ways, that can be represented in the ternary quantum
system as $\ket{000} \dots \ket{222}$. Thus, our database contains 27
elements. In Figure \ref{Fig9}, the gate level representation of the Grover's circuit
is shown. The simulation steps of Figure \ref{Fig9} are:
\par Step 1: At first, all the qutrits of the 3-qutrit
register are initialized with $\ket{0}$.
\par Step 2: Then, the $CH$ gate is applied to create the all possible 27
states $(\ket{000} .... \ket{222})$.
\par Step 3: The ternary oracle compares the color of every
adjacent vertices and inverts the amplitude of solution elements,
which are $\ket{011}, \ket{012}, \ket{100}, \ket{120}, \ket{200}, \ket{201}$ etc.
\par Step 4: The output of ternary oracle is acted upon by the
ternary diffusion operator. This diffusion
operator amplifies the amplitudes of the marked states of step
2.
\par Step 5: Steps 3 and 4 are repeated for 
$\sqrt{N/M}$ times. (For the multiple solution of Grover’s operator having N= number of
elements in the database and M= number of marked states).

\begin{figure}[!htb]
\centering
\includegraphics[width=90mm, height=4.0cm]{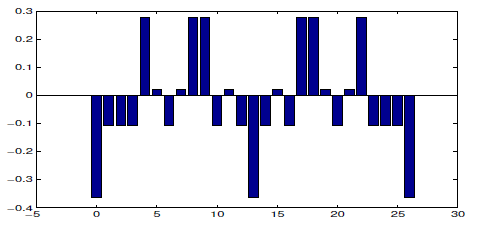}
\caption{Simulation output of oracle after one iteration}
\label{Fig12}
\end{figure}

 Step 5 after one iteration is shown in Figure \ref{Fig12} where, the amplitude amplification is performed using diffusion operator to amplify the amplitudes of marked state. It can be verified that the
location of the searched states (output of step 2) by subtracting
one from the index value of Figure \ref{Fig12} and then converting it
to its equivalent ternary value. For example, if index
value is taken five form Figure \ref{Fig12}, after subtracting one and converting it to
ternary we get the string $\ket{011}$, which is one of the searched
state. Analysis of the simulation result confirms that our oracle successfully verifies that any set of vertices of a graph is properly colored by the given set of colors.

\subsection{Comparative Analysis}

Our proposed comparator-based oracle gives better result with respect to data qubit and ancilla qubit as $n * \lceil \log_{2} k \rceil$ and $ O(n)$ respectively, as compared to \cite{ibm_graphcolor}. The comparative analysis is shown in Table \ref{compare}.


\begin{table}[!htb]
\centering
\caption{Comparative Analysis of Binary Oracle}
\begin{tabular}{ |c|c|c|}
  \hline

Parameters & Hu et. al. \cite{ibm_graphcolor} & Proposed work \\ \hline
Data Qubit Cost & $n*k$ & $n * \lceil \log_{2} k \rceil$\\
 \hline
 Ancilla Qubit Cost & $O((n*k)^2)$ & $ O(n)$\\
 \hline
 Processor & IBMQ & Any NISQ Device\\ \hline
\end{tabular}
\label{compare}
\end{table}

The new ternary
oracle circuit for 3 vetices graph achieved  41\% reduction in the quantum gate cost as
compared to the most recent related work \cite{sbm_graphcolor} and 85\% reduction in the quantum gate cost as compared to the work \cite{wang_color}. Table \ref{ternary_compare} shows the comparative analysis for higher vertices graphs as well. We show that for 4-vertices and 5-vertices graph, it can be reached to 43\% reduction of gate cost. 

\begin{table}[!h]
\centering
\caption{Comparative Analysis of Ternary Oracle}{%
\begin{tabular}{ |l|l|l|l|l|l| }
  \hline

No. of vertex & Proposed Gate Cost & Gate cost\cite{sbm_graphcolor} & Reduction(\%) & Gate cost\cite{wang_color} & Reduction(\%) \\ \hline
$3$ &  $62$ & $106$ & $41\%$ & $343$ & $85\%$ \\ \hline
$4$ &  $170$ & $298$ & $43\%$ & $<1000$ & $86\%$\\ \hline
$5$ & $282$ & $494$ & $43\%$ & $<2700$ & $86\%$\\ \hline
\end{tabular}}
\label{ternary_compare}
\end{table}%

\section{Conclusion}
Here, in this paper, we have proposed an end-to-end framework, which includes mapping of $k$-coloring problem to any NISQ devices/multi-valued quantum technology through automatic generation of oracle circuit using Grover's search followed by MCT realization for any dimensional quantum system. This proposed approach for any dimensional quantum system is applicable for any undirected and unweighted given graph, which makes our approach generalized in nature. Our comparator-based approach can reduce qubit cost to $n * \lceil log_2 {k} \rceil$ as compared to $n * k$ of reduction-based approach from 3-SAT problem to 3-Color problem. This leads to a reduction of query complexity from $O(n * k)$ to $O(n * log_2 {k})$. 
Further, we have shown that our newly proposed ternary comaparator is a key component in designing the circuit for the $k$-coloring problem using quantum search algorithm. Our new ternary
oracle circuit achieved an at least 41\% reduction in the quantum gate cost as
compared to the most recent related work \cite{sbm_graphcolor} and an at least 85\% reduction in the quantum gate cost as compared to the work \cite{wang_color}. In future, with the evolution of qudit-supported quantum hardware, we would like  to validate our designs. 
%


%
%



\end{document}